\documentclass[12pt]{article}

\usepackage{latexsym,amsmath,amssymb,theorem,epsfig}

\DeclareFontFamily{OT1}{rsfs10}{}
\DeclareFontShape{OT1}{rsfs10}{m}{n}{ <-> rsfs10 }{}
\DeclareMathAlphabet{\mathscript}{OT1}{rsfs10}{m}{n}

\numberwithin{equation}{section}

\newcommand{\ns}{\normalsize}

\def\e{\epsilon}

\def\gsim{ \lower .75ex \hbox{$\sim$} \llap{\raise .27ex \hbox{$>$}} }
\def\lsim{ \lower .75ex \hbox{$\sim$} \llap{\raise .27ex \hbox{$<$}} }
\def\be{\begin{equation}}
\def\ee{\end{equation}}
\def\bea{\begin{eqnarray}}
\def\eea{\end{eqnarray}}

\theoremstyle{plain}

{\theorembodyfont{\rmfamily} }

\begin{document}


\begin{titlepage}

\vspace{-5cm}

\title{
  \hfill{\ns }  \\[1em]
   {\LARGE Infrared Limit of Gluon Amplitudes at Strong Coupling}
\\[1em] }
\author{
   Evgeny I. Buchbinder 
     \\[0.5em]
   {\ns Perimeter Institute for Theoretical Physics} \\[-0.4cm]
{\ns Waterloo, Ontario, N2L 2Y5, Canada}}

\date{}

\maketitle

\begin{abstract}

In this note, we propose that the infrared structure
of gluon amplitudes at strong coupling can be fully extracted
from a local consideration near cusps. This is consistent with
field theory and correctly reproduces the infrared divergences 
of the four-gluon amplitude at strong coupling calculated recently 
by Alday and Maldacena. 

\end{abstract}

\thispagestyle{empty}

\end{titlepage}


\section{Introduction}


In the recent paper~\cite{Juan}, Alday and Maldacena made a substantial 
progress in applying ideas of the AdS/CFT correspondence~\cite{Juan1}
to study scattering amplitudes of gluons at strong coupling. 
One of the crucial ingredients of~\cite{Juan} is 
dimensional regularization on the gravity dual side. 
The importance of it is motivated by the fact that
many field theory results on gluon amplitudes have been obtained in dimensional 
regularization scheme and it is necessary to use the same 
regularization if one intends to provide an unambiguous comparison 
between the gravity and field theory sides. 
In particular, Alday in Maldacena~\cite{Juan} computed the 
four-gluon amplitude at strong coupling and found a perfect agreement 
with the infrared structure in field theory. In addition, they 
also found an agreement with the conjecture of Bern, Dixon and Smirnov~\cite{BDS}
(see also an earlier paper~\cite{ABDK})
regarding the all-loop iterative structure of gluon amplitudes. 
Furthermore, the analysis of~\cite{Juan} made a prediction for the 
strong coupling behavior of the cusp anomalous dimension.
Their result agreed with~\cite{GKP}, \cite{Martin} where the same behavior 
was established by different methods. 

It is well-known that in field theory, gluon amplitudes have to satisfy several consistency
conditions such as unitarity cuts, infrared behavior, collinear 
and soft gluon limits. It is very interesting to understand what they are translated 
on the AdS side to. The simplest limit that one can consider is the infrared 
divergences. In field theory they arise from a very special set of Feynman diagrams. 
This makes them summable to all orders in perturbation 
theory~\cite{Sterman1, Catani, Sterman2, BDS}. 
This suggests that on the AdS side the infrared structure also arises
from some very special minimal worldsheets whose study does not 
require the complete answer for the $n$-gluon amplitude. 
In this short note, we propose that the infrared behavior can be completely 
derived from the worldsheet whose boundary is momenta of two neighboring gluons 
meeting at a cusp. We show that this is consistent with the infrared structure 
on the field theory side. When applied to the case of four gluons,
this proposal gives the same answer as the infrared divergent contribution 
to the four-gluon amplitude found in~\cite{Juan}.


\section{Infrared Behavior in ${\cal N}=4$ Super Yang-Mills Theory}


The infrared behavior of the gluon amplitudes in ${\cal N}=4$ 
Super Yang-Mills Theory
is known to all orders in perturbation theory and has a nice exponential 
structure~\cite{Sterman1, Catani, Sterman2, BDS}. In our review below, we will follow 
section IV of~\cite{BDS}. 
We will assume that we have performed the color decomposition and 
study the leading-color partial amplitudes. 
First, let us define 
\begin{equation}
M_n^{(L)}(\e)=\frac{{\cal A}_n^{(L)}(\e)}{{\cal A}_{n}^{tree}}. 
\label{1.1}
\end{equation}
In this equation, ${\cal A}_{n}^{tree}$ and ${\cal A}_n^{(L)}(\e)$ are 
the tree-level and $L$-loop
n-gluon amplitudes respectively. The dependence on $\e$ indicates that $M_n^{(L)}$
is evaluated in dimensional regularization. Furthermore, let us define 
\begin{equation}
M_n(\e)=1+ \sum_{L=1}^{\infty} a^L M_{n}^{(L)}(\e), 
\label{1.2}
\end{equation}
where
\begin{equation}
a = \lambda (4 \pi e^{-\gamma})^{-\e},
\label{1.3}
\end{equation}
$\lambda$ is the t'Hooft coupling and $\gamma$ is the Euler constant. 
For any neighboring pair of gluons $i, i+1$ with momenta $k_i$ and $k_{i+1}$ 
we introduce 
\begin{equation}
s_{i, i+1}=(k_i + k_{i+1})^2, \quad i=1, \dots, n, \quad s_{n, n+1} \equiv s_{n 1}. 
\label{1.4}
\end{equation}
The leading-color all-loop infrared behavior of $M_{n}(\e)$ can be expressed as follows
\begin{equation}
\ln M_{n}(\e) \sim \sum_{l=1}^{\infty} a^l f^{(l)}(\e) \hat{I}_n^{(1)}(l \e). 
\label{1.5}
\end{equation}
Here $\hat{I}_n^{(1)}(\e)$ represents the infrared behavior at one loop and
has the following additive structure
\begin{equation}
\hat{I}_n^{(1)}(\e)= \sum_{i=1}^{n} \hat{I}_n^{(1)}(s_{i, i+1}, \e),
\label{1.6}
\end{equation}
where 
\begin{equation}
\hat{I}_n^{(1)}(s_{i, i+1}, \e) = -\frac{1}{2 \e^2} 
\left( \frac{\mu^2}{-s_{i, i+1}}\right)^{\e}. 
\label{1.61}
\end{equation}
The function $f^{(l)}(\e)$ has a perturbative expansion in $\e$ 
\begin{equation}
f^{(l)}(\e)=f_0^{(l)}+ f_1^{(l)}\e +f_2^{(l)}\e^2 + \dots.
\label{1.7}
\end{equation}
The leading term $f_0^{(l)}$ is known to coincide, up to a constant, 
with the cusp anomalous dimension $\gamma^{(l)}$, 
\begin{equation}
f_0^{(l)}=\frac{1}{4}\gamma^{(l)}. 
\label{1.8}
\end{equation}
Substituting $\hat{I}_n^{(1)}(\e)$ into eq.~\eqref{1.5} we obtain that 
the right hand side has the following functional dependence
\begin{equation}
\ln M_n(\e) \sim \frac{1}{\e^2}\sum_{i=1}^{n} 
F \left(\lambda \left(\frac{\mu^2}{-s_{i, i+1}}\right)^{\e}, \e \right)
\label{1.9}
\end{equation}
for some function $F$ which goes to zero for finite negative $\e$ as 
$s_{i, i+1}$ goes to zero. 
To find another implication of the general infrared behavior~\eqref{1.5}
let us consider the terms of order $\e^0$. It follows from 
eqs.~\eqref{1.6}, \eqref{1.61} and~\eqref{1.7} that up to an additive constant 
$\ln M_n(\e)$ contains the the terms of the form
\begin{equation}
\ln M_n(\e)|_{\e^0} \sim -\frac{1}{16}f(\lambda) \sum_{i=1}^{n}
\ln^2 \left( \frac{\mu^2}{-s_{i, i+1}}\right) 
- \frac{1}{4}g(\lambda) \sum_{i=1}^{n}
\ln \left( \frac{\mu^2}{-s_{i, i+1}}\right), 
\label{1.10}
\end{equation}
where
\begin{equation}
f(\lambda) =4\sum_{l=1}^{\infty} \lambda^l f_0^{(l)}=\sum_{l=1}^{\infty}\lambda^l \gamma^{(l)}
\label{1.11}
\end{equation}
is the all-loop cusp anomalous dimension and
\begin{equation}
g(\lambda) =2\sum_{l=1}^{\infty} \frac{\lambda^l f_1^{(l)}}{l}-8 \pi e^{-\gamma}\sum_{l=1}^{\infty}f_0^{(l)}. 
\label{1.12}
\end{equation}
Note that the definition of $g(\lambda)$ depends on the infrared scale $\mu$. 
As we change $\mu$, $\mu \to \mu \kappa$, we have 
\begin{equation}
g(\lambda) \to g(\lambda) + 2 f(\lambda) \ln \kappa. 
\label{1.13}
\end{equation}
On the other hand, the coefficient in front of the $\ln^2 \left( \frac{\mu^2}{-s_{i, i+1}}\right)$
term is always the cusp anomalous dimension.


\section{Infrared Behavior and Cusps in AdS}


In~\cite{Juan}, Alday and Maldacena developed an approach to study the 
strong coupling limit of ${\cal N}=4$ gluon amplitudes using 
string theory in AdS space. 
As explained in~\cite{Juan}, scattering of open string states 
happens at large proper AdS momenta. Therefore, similarly to the flat space
case~\cite{Gross1}, the scattering amplitude to the leading order 
is determined by the appropriate classical solution
\begin{equation}
{\cal A}\sim e^{ i S}, 
\label{3.2}
\end{equation}
where $S$ is the action evaluated on the classical solution, which is 
just the area of the worldsheet. 
The prefactor was determined in~\cite{Khoze}.
Furthermore, similarly to the 
flat space case~\cite{Gross1}, ${\cal A}$ depends only on the momenta of the scattered
particles and not on any other additional data like the helicity structure. All 
information about the momenta is encoded in the boundary conditions. To describe it, it
is convenient to use the T-dual coordinates~\cite{KT} (see~\cite{Juan} for details). 
\begin{figure}[ht]
\centering
\includegraphics[width=70mm]{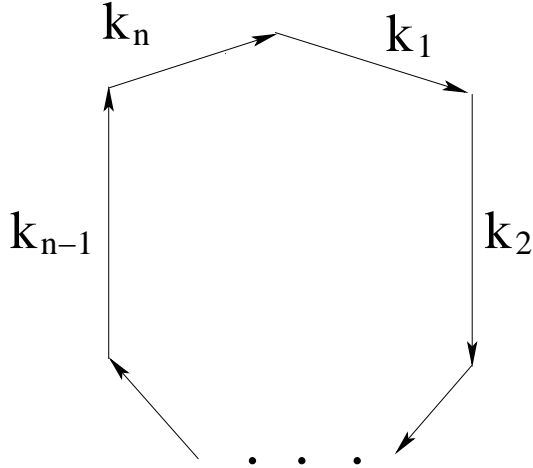}
\caption{The boundary of the worldsheet of the $n$-gluon amplitude.
For simplicity, we removed the factors of $2 \pi$ multiplying each $k_{i}$.}
\label{pot1}
\end{figure}
In this coordinates, the metric is also AdS 
\begin{equation}
ds^2= R^2 \frac{dy_{\mu} dy^{\mu} +dr^2}{r^2}, \quad \mu =0, \dots, 3, 
\label{3.3}
\end{equation}
where $R$ is the radius of AdS space. The boundary conditions are imposed at $r=0$. 
In the T-dual coordinates the fact that a state has momentum $k^{\mu}$ translates 
into the statement that it has a winding 
\begin{equation}
\Delta y^{\mu} =2 \pi k^{\mu}. 
\label{3.5}
\end{equation}
Then, the boundary conditions are such that as $r \to 0$ the worldsheet 
describing the $n$-particle amplitude ends on the vectors
$2 \pi k_1^{\mu}, \dots, 2 \pi k_n^{\mu}$. The ordering of the vectors corresponds 
to the particular color ordered amplitude. From momentum conservation it follows that 
the above vectors form a closed loop. 

Since ${\cal N}=4$ gluon amplitudes are infrared divergent they have to be regulated. 
In order to be able to compare the string and field theory results, one has to use 
the same regularization scheme. The most convenient is to use dimensional regularization.
The regulated AdS metric looks as follows~\cite{Juan}
\begin{equation}
d s^2 = \sqrt{c_D \lambda_D} \left(\frac{dy^2_D + dr^2}{r^{2+\epsilon}} \right), 
\label{3.6}
\end{equation}
where 
\begin{eqnarray}
&&\lambda_D= \frac{\lambda \mu^{2 \epsilon}}{(4 \pi e^{-\gamma})^{\epsilon}}, \quad
c_D=2^{4 \epsilon} \pi^{3 \epsilon} \Gamma (2+\epsilon), \nonumber \\
&&D= 4-2 \e. 
\label{3.7}
\end{eqnarray}
The parametrization of $\lambda_D$ in terms of the IR scale $\mu$ is chosen 
to match the field theory side. In notation~\eqref{3.6}, the worldsheet action becomes
\begin{figure}[ht]
\centering
\includegraphics[width=150mm]{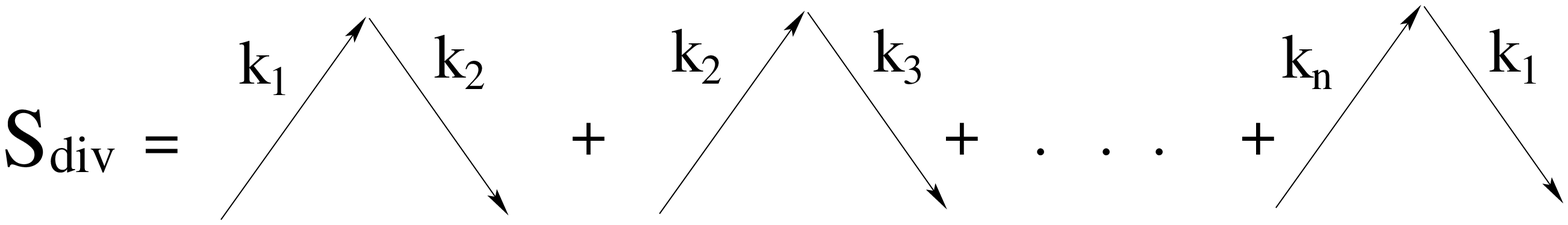}
\caption{The graphical representation of the infrared divergences.
Each cusp represents the boundary of the minimal worldsheet.}
\label{pot2}
\end{figure}
\begin{equation}
S=\frac{\sqrt{\lambda_D c_D}}{2 \pi} \int \frac{{\cal L}_{\e=0}}{r^{\e}}, 
\label{3.8}
\end{equation}
where ${\cal L}_{\e=0}$ is the Lagrangian evaluated at $\e=0$, that is using the 
metric~\eqref{3.3} where the AdS radius is set to unity. 

Let us now consider the boundary of the worldsheet corresponding to the $n$-gluon 
amplitude as in Figure 1. 
The boundary consists of $n$ vectors 
$2 \pi k_1^{\mu}, \dots, 2 \pi k_n^{\mu}$. 
We will denote by 
${\cal C}_i$ the cusp  where the vectors $2 \pi k_i^{\mu}$ and
$2 \pi k_{i+1}^{\mu}$ meet. In this note, we propose that the infrared 
behavior of the $n$-gluon amplitude at strong coupling is fully captured 
by the local behavior of the worldsheet near the $n$ cusps. To be more precise, 
we propose that the infrared divergences have the structure as in Figure 2.
That is,~\footnote{In eq.~\eqref{3.8.1}, it does not matter 
whether we write $\ln {\cal A}_{n}$ or $\ln M_n$ since 
the difference between them, $\ln {\cal A}_n^{tree}$ 
is independent of $\lambda$ and, thus, is subleading at strong coupling.}
\begin{equation}
\ln M_{n} \sim i S_{div}(\e) =\sum_{i=1}^n i S_{i, i+1}(\e).
\label{3.8.1}
\end{equation}
The summation is over all the pairs of neighboring gluons or, 
equivalently, over all cusps. 
The $i$-th term in the right hand sides in eq.~\eqref{3.8.1} and in  
Figure 2 represents the area $S_{i, i+1}(\e)$
of the minimal worldsheet 
whose boundary is just the two vectors $2 \pi k_i^{\mu}$ and
$2 \pi k_{i+1}^{\mu}$. As a trivial consistency check, we note 
that eq.~\eqref{3.8.1}
has the same additive structure as eqs.~\eqref{1.5}, \eqref{1.6}
on the field theory side. 

Since eq.~\eqref{3.8.1} and Figure 2 have an additive structure, 
it is sufficient to single out only one cusp ${\cal C}_i$ and consider 
the problem of finding a minimal worldsheet which ends on the vectors
$2 \pi k_i^{\mu}$ and $2 \pi k_{i+1}^{\mu}$. 
We want to point out that our calculation is 
universal and does not depend on global structure of the amplitude. 
Without loss of generality, we can assume that the worldsheet is located 
in the subspace parametrized by $(y_0, y_1, y_2, r)$ and set 
$y_3=0$.~\footnote{In principle, one can choose a coordinate system in which 
two light-like vectors lie in a two-plane. The choice of the coordinate system is a matter of convenience.
The physical conclusions studied below are, of course, independent of this choice.}
It is convenient to introduce the light-cone coordinates in the 
$(y_0, y_1)$ plane
\begin{equation}
y_-= y_0 -y_1, \quad y_+= y_0+y_1. 
\label{3.18}
\end{equation}
The metric in these coordinates becomes
\begin{equation}
ds^2=\frac{1}{r^{2+\e}}(-dy_+ dy_- +dy_2^2 +dr^2). 
\label{3.19}
\end{equation}
The action is given by eq.~\eqref{3.8}
\begin{equation}
S_{i, i+1}(\e) =\frac{\sqrt{\lambda_D c_D}}{2 \pi} \int \frac{{\cal L}_{\e=0}}{r^{\e}}. 
\label{3.20}
\end{equation}
We will parametrize the worldsheet by $(y_-, y_+)$. Then it follows that 
\begin{equation}
{\cal L}_{\e=0}=dy_- dy_+ \frac{1}{2 r^2}
\sqrt{1-4 \partial_- y_2  \partial_+ y_2
-4 \partial_- r  \partial_+ r -
4 (\partial_- y_2  \partial_+r -\partial_- r  \partial_+ y_2)^2}. 
\label{3.21}
\end{equation}
We need to find a solution that turns into two intersecting lines 
whose directions are specified by $k_i$ and $k_{i+1}$
as $r \to 0$. 
We can choose the coordinate system in such a way that 
one of the vectors, say $k_i$, is located in the $(y_-, y_+)$ plane. 
Moreover, we can chose $k_i$ to lie along the $y_+$ direction.
We parametrize
it as
\begin{equation}
2 \pi k_i=z_1 (0, 1, 0)
\label{3.22}
\end{equation}
in the $(y_-, y_+, y_2)$ coordinates. The parameter $z_1$ is arbitrary as long as 
it is non-zero and finite. 
Similarly, we parametrize 
the vector $k_{i+1}$ as
\begin{equation}
2 \pi k_{i+1}=z_2 (\alpha, 1, \sqrt{\alpha}),
\label{3.23}
\end{equation}
where $\alpha$ is the tangent of the angle between the lines when they are projected 
to the $(y_-, y_+)$ plane and $z_2$, like $z_1$, is an arbitrary non-zero finite 
parameter.From eqs.~\eqref{3.22}
and~\eqref{3.23} it follows that 
\begin{equation}
(2 \pi)^2 s_{i, i+1}= -\alpha z_1 z_2. 
\label{3.24}
\end{equation}
In order to find the solution of interest, we start with the solution ending 
on the lines
\begin{equation}
y_-=0, \quad y_+=0, 
\label{3.25}
\end{equation}
This solution was found in~\cite{Juan}. It looks like
\begin{equation}
r(y_-, y_+)=\sqrt{1+\e/2}\sqrt{2}\sqrt{y_- y_+}, \quad y_2(y_-, y_+)=0. 
\label{3.26}
\end{equation}
The solution we are interested in has to approach~\eqref{3.26} in the limit 
$\alpha \to \infty$. A simple ansatz that satisfies the correct boundary conditions 
and approaches eq.~\eqref{3.26} in the large $\alpha$ limit is 
\begin{equation}
r(y_-, y_+)=\sqrt{1+\e/2}\sqrt{2}\sqrt{y_- (y_+ -\frac{1}{\alpha} y_-)}, 
\quad y_2(y_-, y_+)=\frac{1}{\sqrt{\alpha}}y_-. 
\label{3.29}
\end{equation}
The ansatz for $y_2$ in eq.~\eqref{3.29} is chosen to make 
sure that the both lines are light-like.
It turns out that eq.~\eqref{3.29} is the exact solutions of the equations of motion 
to all orders in $\e$.
To continue, it is convenient to change the variables from 
$(y_-, y_+)$ to $(Y_-, Y_+)$ so that the $(Y_-, Y_+)$ components 
of $2 \pi k_i$ and  $2 \pi k_{i+1}$ become $(0, 1)$ and $(1, 0)$. The transformation is 
the following 
\begin{equation}
y_-=\alpha z_2 Y_-, \quad
y_+= z_1 Y_+ + z_2 Y_-. 
\label{3.30}
\end{equation}
In these new variables we get 
\begin{equation}
r(Y_-, Y_+)=\sqrt{1+\e/2}\sqrt{2}\sqrt{Y_- Y_+}\sqrt{-(2 \pi)^2 s_{i, i+1}}, 
\quad y_2(Y_-, Y_+)= \sqrt{\alpha} z_2 Y_-. 
\label{3.31}
\end{equation}
Furthermore, 
\begin{equation}
d y_- d y_+= (-(2 \pi)^2 s_{i, i+1})d Y_- d Y_+. 
\label{3.32}
\end{equation}
Substituting eqs.~\eqref{3.30}-\eqref{3.32} into the action~\eqref{3.20}, \eqref{3.21}, 
we obtain
\begin{equation}
i S_{i, i+1} (\e)=-
\frac{\sqrt{\lambda_D c_D}}{4 \pi} \frac{\sqrt{1+\e}}{(1+\e/2)^{1+\e/2}}
(-(2 \pi)^2 s_{i, i+1})^{-\e/2} \int_0^1 
\frac{d Y_- d Y_+}{(2 Y_- Y_+)^{1+\e/2}}. 
\label{3.33}
\end{equation}
Assuming that $\e <0$ and performing the integral, we find that 
\begin{equation}
i S_{i, i+1}(\e)= -\frac{1}{\e^2} \frac{\sqrt{\lambda}}{2 \pi}
\sqrt{\frac{\mu^{2 \e}}{(-s_{i, i+1})^{\e}}} C(\e), 
\label{3.34}
\end{equation}
where $C(\e)$ is given by 
\begin{equation}
C(\e) =\frac{\sqrt{c_D}}{2^{\e/2}} \frac{(2 \pi)^{-\e}}{(4 \pi e^{-\gamma})^{\e/2}}
\frac{\sqrt{1+\e}}{(1+\e/2)^{1+\e/2}}. 
\label{3.35}
\end{equation}
Eq.~\eqref{3.34} represents our final answer for $S_{i, i+1}(\e)$ to all 
orders in $\e$. Note that it depends only on the kinematic invariant
$s_{i, i+1}$ and not separately on $\alpha, z_1$ and $z_2$.
Furthermore, note that eq.~\eqref{3.34} is consistent with 
the general properties of the infrared behavior of ${\cal N}=4$ 
gluon amplitudes reviewed in the previous section. It is of the 
form~\eqref{1.9}, where the function 
$F \left(\lambda \left(\frac{\mu^2}{-s_{i, i+1}}\right)^{\e}, \e \right)$ 
is given by 
\begin{equation}
F \left(\lambda \left(\frac{\mu^2}{-s_{i, i+1}}\right)^{\e}, \e \right)=
-\frac{1}{2 \pi}
\sqrt{\lambda\left( \frac{ \mu^2}{(-s_{i, i+1})}\right)^{\e}} C(\e). 
\label{3.36}
\end{equation}
To compare eq.~\eqref{3.34} with eq.~\eqref{1.10} on the field theory 
side we have to expand $C(\e)$ to the linear order in $\e$. Up to a constant, this 
yields the following $\e$-independent term in $i S_{i, i+1}(\e)$
\begin{equation}
-\frac{1}{16}f(\lambda)
\ln^2 \left( \frac{\mu^2}{-s_{i, i+1}}\right) 
- \frac{1}{4}g(\lambda) 
\ln \left( \frac{\mu^2}{-s_{i, i+1}}\right),
\label{3.37}
\end{equation}
where
\begin{equation}
f(\lambda)=\frac{\sqrt{\lambda}}{\pi}
\label{3.38}
\end{equation}
and 
\begin{equation}
g(\lambda)=\frac{\sqrt{\lambda}}{2\pi}(1 - \ln 2).
\label{3.39}
\end{equation}
Since the general structure of the infrared divergences
implies that the coefficient at 
$\ln^2 \left( \frac{\mu^2}{-s_{i, i+1}}\right)$
is the cusp anomalous dimension, we find that it behaves as
$\sqrt{\lambda}$ at strong coupling which is in agreement 
with~\cite{GKP, Martin, Juan}.

As the last consistency check, let us compare~\eqref{3.34}
with the infrared divergent contribution to the strong coupling limit of 
the four-gluon amplitude
which was obtain by Alday and Maldacena in~\cite{Juan}. 
In the case of the four-gluon amplitude we have only two independent 
kinematic invariants which are usually denoted by $s$ and $t$, 
\begin{equation}
s=s_{1 2}=s_{3 4}, \quad t=s_{2 3}=s_{4 1}. 
\label{3.40}
\end{equation}
As the result, eq.~\eqref{3.8.1} becomes
\begin{equation}
 S_{div}(\e) =2 S_{s}(\e)+2 S_{t}(\e), 
\label{3.41}
\end{equation}
where $S_{s}(\e)$ is given by 
\begin{equation}
i S_{s}(\e)=
-\frac{1}{\e^2} \frac{\sqrt{\lambda}}{2 \pi}
\sqrt{\frac{\mu^{2 \e}}{(-s)^{\e}}} -
\frac{1}{\e}  \frac{\sqrt{\lambda}}{4 \pi}(1-\ln 2) \sqrt{\frac{\mu^{2 \e}}{(-s)^{\e}}}
+{\cal O}(\e^0) 
\label{3.42}
\end{equation}
and $S_{t}(\e)$ is given by the similar expression with $s$ replaced with $t$. 
Quite remarkably, eqs.~\eqref{3.41} and~\eqref{3.42}
exactly coincide with the infrared behavior of the four-gluon 
amplitude computed in~\cite{Juan} which represents a non-trivial check of our proposal. 

To conclude, we have proposed that the infrared structure of the 
gluon amplitudes at strong coupling is fully captured
by the behavior near the cusps as indicated in Figure 2. 
The total $n$-gluon amplitude is then given by 
\begin{equation}
\ln M_n = \sum_{i=1}^n i S_{i, i+1}(\e) +{\cal F}_n, 
\label{3.43}
\end{equation}
where $i S_{i, i+1}(\e)$ is given by eq.~\eqref{3.34} and
${\cal F}_n$ is the $\mu$-independent finite contribution which is often reffered to as 
the finite remainder. It is independent of the helicity 
structure and, thus, is a function only of the kinematic invariants. 
The calculation of ${\cal F}_n$ is, certainly, a very complicated 
problem and one can expect that understanding of unitarity cuts and collinear behaviour 
in AdS can provide a significant help.


\section{Acknowledgements}


The author is very grateful to Freddy Cachazo for helpful discussions. 
The author would also like to thank Juan Maldacena and Arkady Tseytlin for comments on the first 
version of the paper.
The work at the Perimeter Institute is supported by in part by the Government of Canada through NSERC 
and by the Province of Ontario through MRI.




\begin{thebibliography}{99}


\bibitem{Juan}
L. F. Alday and J. Maldacena, 
``Gluon scattering amplitudes at strong coupling,''
arXiv:0705.0303.

\bibitem{Juan1}
J. M. Maldacena, 
``The Large N Limit of Superconformal Field Theories and Supergravity,''
Adv.Theor.Math.Phys. 2 (1998) 231-252; Int.J.Theor.Phys. 38 (1999) 1113-1133
[arXiv:hep-th/9711200].

\bibitem{BDS}
Z. Bern, L. J. Dixon and V. A. Smirnov, 
``Iteration of Planar Amplitudes in Maximally Supersymmetric Yang-Mills Theory at 
Three Loops and Beyond,''
Phys.Rev. D72 (2005) 085001 [arXiv:hep-th/0505205].

\bibitem{ABDK}
C.~Anastasiou, Z.~Bern, L.~Dixon and D.~A.~Kosower,
``Planar Amplitudes in Maximally Supersymmetric Yang-Mills Theory,''
Phys.Rev.Lett. 91 (2003) 251602 [hep-th/0309040].

\bibitem{GKP}
S.~S.~Gubser, I.~R.~Klebanov and A.~M.~Polyakov, 
``A semi-classical limit of the gauge/string correspondence,''
Nucl.Phys. B636 (2002) 99-114 [arXiv:hep-th/0204051]. 

\bibitem{Martin}
M.~Kruczenski, 
``A note on twist two operators in N=4 SYM and Wilson loops in Minkowski signature,''
JHEP 0212 (2002) 024 [arXiv:hep-th/0210115].

\bibitem{Sterman1}
L. Magnea and G. Sterman, 
``Analytic continuation of the Sudakov form-factor in QCD,''
Phys.Rev.D42:4222-4227,1990. 

\bibitem{Catani}
S. Catani, 
``The Singular Behaviour of QCD Amplitudes at Two-loop Order,''
Phys.Lett. B427 (1998) 161-171 [arXiv:hep-ph/9802439].

\bibitem{Sterman2}
G. Sterman and M. E. Tejeda-Yeomans, 
``Multi-loop Amplitudes and Resummation,''
Phys.Lett. B552 (2003) 48-56 [arXiv:hep-ph/0210130].
 
\bibitem{Gross1} 
D.~J.~Gross and P.~F.~Mende, 
`` The High-Energy Behavior of String Scattering Amplitudes,''
Phys.Lett.B197 (1987) 129.

\bibitem{Khoze}
S. Abel, S. Forste and V. V. Khoze, 
``Scattering amplitudes in strongly coupled N=4 SYM from semiclassical strings in AdS,''
[arXiv:0705.2113].

\bibitem{KT}
R.~Kallosh and A.~A.~Tseytlin,
``Simplifying superstring action on $AdS_5 \times S^5$,''
JHEP 9810 (1998) 016 [arXiv:hep-th/9808088].


\end{thebibliography}
\end{document}